\documentclass[namedreferences,hyperref,optionalrh]{spr-sola}
\usepackage{graphicx}        
\usepackage{color}           

\chardef\us=`\_

\newcommand{\Ly}{Ly$\alpha\;$}
\newcommand{\flrthree}{SOL2012-07-04}

\newcommand{\flrfour}{SOL2009-12-10}

\newcommand{\comment}{}
\newcommand{\commentend}{}
\usepackage{tabularx} 
\usepackage{makecell} 
\usepackage[T1]{fontenc} 
\usepackage{anyfontsize} 
\usepackage{booktabs} 
\usepackage{rotating} 
\usepackage{amssymb}

\mathchardef\mhyphen="2D

\graphicspath{{Figures_Mockup}}

\usepackage{geometry}
\begin{document}

\begin{article}
\begin{opening}
\title{Observations of Flare Induced Doppler Shifts in the Si~\textsc{iii} $1206\,\textrm{Å}$ line}

\author[addressref=aff1,corref,email={lmajury01@qub.ac.uk}]{\inits{L.H.}\fnm{Luke}~\lnm{Majury}\orcid{0009-0002-8491-9593}}

\author[addressref=aff1,corref,email={r.milligan@qub.ac.uk}]{\inits{R.O.}\fnm{Ryan}~\lnm{Milligan}\orcid{0000-0001-5031-1892}}

\address[id=aff1]{Astrophysics Research Centre, School of Mathematics and Physics, Queen's University Belfast, University Road, BT7 1NN, Northern Ireland, UK}


\runningauthor{Majury et al.}
\runningtitle{Observations of Chromospheric Condensation in the Si~\textsc{iii} 1206 Å line}

\begin{abstract}
    
    Doppler shifts in chromospheric and transition-region lines during solar flares are often interpreted as chromospheric condensation or evaporation. However, alternative sources of Doppler-shifted emission have been suggested, such as filament eruptions, jets or chromospheric bubbles. We analyse high-cadence scans from SORCE/SOLSTICE, which provide one-minute resolution profiles of the transition-region Si~\textsc{iii} ($1206\,\textrm{Å}$, $\textrm{T} = 10^{4.6}\,\textrm{K}$) line. 11 X-, M-, and C-class events observed during these scans with clear impulsive phase Si~\textsc{iii} enhancements were identified. By subtracting a quiet-Sun profile and fitting Gaussian profiles to the Si~\textsc{iii} line, measurements of flare-induced Doppler shifts were made. After correcting for a systematic trend in these shifts with solar longitude, two of the 11 events were found to exhibit a significant Doppler shift, \comment{}one with a $201.36\pm21.94\;\textrm{km\,s}^{-1}$ redshift and the other with a $-39.75\pm11.00\;\textrm{km\,s}^{-1}$ blueshift\commentend{}. Intriguingly, SDO/AIA $304\,\textrm{Å}$ and $1600\,\textrm{Å}$ imaging revealed a bright eruption coincident with the event that exhibited a blueshift, suggesting the shift may have resulted from the eruption rather than evaporation alone. Our results highlight Si~\textsc{iii} as a useful diagnostic of flaring dynamics at a temperature that has received limited attention to date. Future comparisons of these observations with radiative hydrodynamic simulations, along with new observations from state-of-the-art spectrometers such as SOLAR-C/EUVST and MUSE, should clarify the mechanisms behind the observed shifts in this study.
    
\end{abstract}

\end{opening}
\section{Introduction}
    Chromospheric evaporation and condensation are fundamental processes in the deposition and transport of flare energy throughout the solar atmosphere. Evaporation generally results from impulsive heating of the solar chromosphere, with heated ambient material ablating to the solar corona, filling the flare loops and emitting strongly in Soft X-rays (SXR) and Extreme Ultraviolet (EUV) radiation. The gentle evaporation process is inferred through blue Doppler shifts on the order of $10\;\textrm{km}\,\textrm{s}^{-1}$ for lines at temperatures between $10^4$ and $10^6\,\textrm{K}$, with relatively faster velocities, often exceeding $100\;\textrm{km}\,\textrm{s}^{-1}$, for hotter lines \citep{Milligan_2006_2,delZanna_2006_2}. For strong heating rates of the chromosphere ($>3\times10^{10}\;\textrm{erg}\,\textrm{s}^{-1}\,\textrm{cm}^{-2}$), explosive evaporation occurs, driving fast upflows at several hundreds of $\textrm{km}\,\textrm{s}^{-1}$ that are observed in lines with $\textrm{T} \gtrsim 10^6$ \citep{Fisher_1985_1,Fisher_1985_2,Fisher_1985_3}. With such heating rates, chromospheric condensation occurs, with strong overpressure driving downflows at velocities of tens of $\textrm{km}\,\textrm{s}^{-1}$ in the chromosphere and transition region that balance momentum with the upflowing material \citep{Milligan_2006_1,delZanna_2006_1,Milligan_2009}. These processes provide vital insight into the transport of flare energy throughout the solar atmosphere, linking impulsive heating and energy release to gradual phase emission during flares.
    
    Historically, Doppler shifts in chromospheric lines have been frequently attributed to chromospheric evaporation. However, more recently, other possibilities for these shifts have been explored. \citet{Batchelor_1991} suggested that a blueshift observed in the coronal Ca~\textsc{xix} line during a flare was driven by emission from an erupting filament. A similar blueshift in the chromospheric H$\alpha$ line was attributed to an erupting filament during a flare by \citet{Ding_2003}. In \citet{Brown_2016}, flares observed by the Extreme ultraviolet Variability Experiment on the Solar Dynamics Observatory \citep[SDO/EVE;][]{Woods_2012,Pesnell_2012} that exhibited blueshifts in chromospheric lines were found to be associated with eruptions, which potentially drove the shifts. Furthermore, \citet{Tei_2018} discussed the possibility that a chromospheric bubble drove the blueshift observed in the Mg~\textsc{ii} h line during a flare observed by the Interface Region Imaging Spectrometer \citep[IRIS;][]{DePontieu_2014}. This blueshift transitioned to a redshift in the later phases of the flare, consistent with chromospheric condensation. Blueshifted chromospheric emission during flares may also be associated with jets. \citet{Zhang_2021} studied a jet associated with a C3.4 flare, reporting a blueshift ranging from $-34\;\textrm{km}\,\textrm{s}^{-1}$ to $-120\;\textrm{km}\,\textrm{s}^{-1}$ in the Si~\textsc{iv} $1403\,\textrm{Å}$ line observed by IRIS. However, the ribbons showed positive Doppler velocities in the line, indicating chromospheric condensation.\comment{} Apparent motions during flares can also be driven by enhancement of line blends, making identification of such blends important to ensure accurate detections of Doppler shifts \citep[e.g.][]{Li_2016}\commentend{}.

    Further insights into flare-driven Doppler shifts in chromospheric lines could be gained by studying chromospheric and transition-region lines that probe scarcely explored temperatures. In particular, comparisons with well-studied lines may clarify how flare dynamics influence Doppler signatures and how the formation heights of different lines change during flares. Recently publicised wavelength calibration scans of the far ultraviolet (FUV) by the Solar-Stellar Irradiance Comparison Experiment on board the Solar Radiation and Climate Experiment \citep[SORCE/SOLSTICE\footnote{https://lasp.colorado.edu/sorce/data/};][]{Rottman_2005,Mcclintock_2005} offer an opportunity to study Doppler shifts in the previously unstudied Si~\textsc{iii} line \citep[$1206\,\textrm{Å}$, $\textrm{T} = 10^{4.6}\,\textrm{K}$; CHIANTI\footnote{https://www.chiantidatabase.org/chianti\_linelist.html} Version 11.0;][]{Dere_1997,Dufresne_2024} and ascertain the physical mechanism driving them. The temperature of the line bridges between previously observed lines at this temperature, such as the C~\textsc{ii} ($1336\,\textrm{Å}$, $\textrm{T} = 10^{4.5}\,\textrm{K}$) and  Si~\textsc{iv} ($1403\,\textrm{Å}$, $\textrm{T} = 10^{4.8}\,\textrm{K}$) lines observed by IRIS, potentially allowing unique insights to be yielded. 
    
    IRIS observations of different lines of the same atomic species as Si~\textsc{iii}, including the aforementioned Si~\textsc{iv} $1403\,\textrm{Å}$ line and several Si~\textsc{ii} lines, show evidence for chromospheric condensation in flare ribbons \citep[e.g.][]{Tian_2015,Graham_2020,Zhang_2021,Lorincik_2022_2}. However, little is known of the behaviour of the Si~\textsc{iii} line during flares. Past studies using SORCE/SOLSTICE wavelength calibration scans note strong relative and absolute enhancement of the line that occurs predominantly during the impulsive phase of flares, properties that may make the line a useful diagnostic of line shifts in the chromosphere \citep{Woods_2004, Milligan_2016, Majury_2025, Li_2025}.
    
    In this paper, we present Doppler velocities derived from shifts in the Si~\textsc{iii} line during the impulsive phase of 11~X-, M- and C-class flares observed in disk-integrated SORCE/SOLSTICE wavelength calibration scans. Section \ref{S- Observations} describes the SORCE/SOLSTICE wavelength calibration scans and the methods used to derive Doppler velocities from the Si~\textsc{iii} profile for each flare. Section \ref{S- Results} presents the Doppler velocities calculated for each event along with lightcurves of two flares, which had coverage in ultraviolet (UV) and X-ray emission. Context imaging of chromospheric emissions is briefly analysed for one of these events. Section \ref{S- Discussion} provides interpretation of the flare dynamics behind these results and discusses future avenues for research into the Si~\textsc{iii} line along with other chromospheric and transition-region lines in the context of chromospheric evaporation, condensation, and other processes that may potentially drive Doppler shifts in these lines.
    
\section{Observations  and Analysis}
    \label{S- Observations}
        
\begin{figure}[ht] 
        \centerline{\hspace*{0.015\paperwidth}
               \includegraphics[width=.45\paperwidth,clip=]{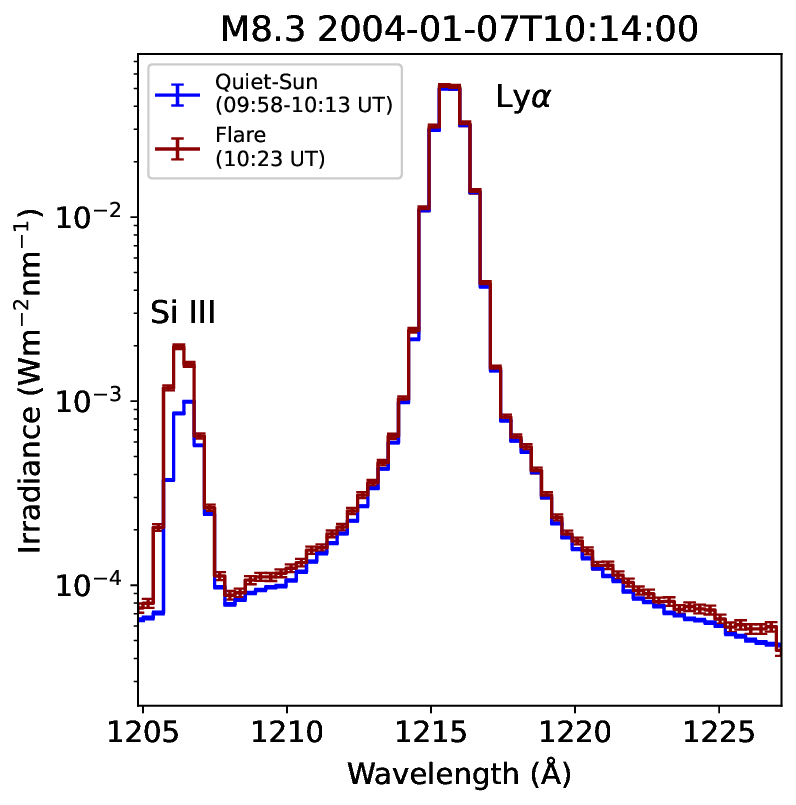}
              }
    \caption{Plot of Si~\textsc{iii} and \Ly line profiles from SORCE/SOLSTICE during the impulsive phase (flare; red) and postflare (quiet-Sun; blue) of M8.3 flare on 7 Jan 2004. Irradiance is plotted on a log scale.}
    \label{figure5.05}
    \end{figure}

    SORCE/SOLSTICE primarily took low-cadence spectral observations of solar irradiance, providing insight into long-term variability in UV wavelengths. However, the instrument also performed high-cadence wavelength calibration scans of the FUV between $1203\,\textrm{Å}$ and $1227\,\textrm{Å}$ for $50$ minutes, approximately every two days of the mission. These observations captured both the \Ly line at $1216\,\textrm{Å}$ and Si~\textsc{iii} line at $1206\,\textrm{Å}$ in raster scans taken in steps of $0.035\,\textrm{Å}$, with each step lasting $1.05\,\textrm{s}$. Further details on these scans are provided in \citet{Mcclintock_2005}, \citet{Snow_2022}, and \citet{Majury_2025}. Each scan across the Si~\textsc{iii} profile took approximately $10\,\textrm{s}$, with the scans being spaced out by the full scan length of $67\,\textrm{s}$ for flares up to 2005. In 2005, the scanning mode of SOLSTICE changed, switching between scanning in positive and negative wavelength directions with each successive scan. This results in observations after 2005 taking two repeated scans of the Si~\textsc{iii} line back to back, with a two-minute gap between these sets of scans. Individual spectral irradiance measurements in each scan have typical uncertainties of $\sim5\%$. Exemplary Si~\textsc{iii} profiles from an M8.3 flare are plotted in Figure \ref{figure5.05}. While SORCE/SOLSTICE provided excellent stability in its wavelength scale at $\sim1\,\textrm{pm}$, spacecraft roll manoeuvres, which were performed a few times per orbit, may shift the absolute wavelength values by several $\textrm{pm}$. These manoeuvres are identifiable as $\sim1^{\prime}$ shifts in the SORCE Fine Sun Sensor modules' (FSS) pointing (Snow, M., Personal Communication, 2025). 

    To support Doppler measurements derived from SORCE/SOLSTICE observations, comparisons are drawn with photometric observations from the E-channel of the Extreme Ultraviolet Sensor on the Geostationary Operational Environmental Satellites \citep[GOES/EUVS-E;][]{Viereck_2007,Evans_2010}, imaging observations from the chromospheric channels of the Atmospheric Imaging Assembly on SDO \citep[SDO/AIA;][]{Lemen_2012}, along with X-ray observations from The Reuven Ramaty High Energy Solar Spectroscopic Imager \citep[RHESSI;][]{Lin_2002} and the X-ray Sensor on GOES (GOES/XRS). This provides further insights into the dynamic processes driving enhancement and Doppler shifts in the Si~\textsc{iii} line. 
    
    EUVS-E provided broadband solar irradiance observations taken at a cadence of $10.24\,\textrm{s}$ with a bandpass between $1180-1270\,\textrm{Å}$, covering both the \Ly and Si~\textsc{iii} lines. The instrument has previously been used to study flares in several works, yielding insights into the timing and energetic significance of chromospheric flare emission \citep{Milligan_2020,Lu_2021,Milligan_2021,Greatorex_2023}. AIA provides chromospheric imaging via its $304\,\textrm{Å}$ and $1600\,\textrm{Å}$ channels which operate at cadences of $12\,\textrm{s}$ and $24\,\textrm{s}$, respectively. The $304\,\textrm{Å}$ channel is dominated by the He~\textsc{ii} resonance line ($\textrm{T} = 10^{4.9}\,\textrm{K}$), which originates primarily from the chromosphere and transition region. The broadband $1600\,\textrm{Å}$ channel captures several emission lines, including the strong C~\textsc{iv} doublet at $1550\,\textrm{Å}$ ($\textrm{T} = 10^{4.95}\,\textrm{K}$), and is dominated by chromospheric emission \citep{Simoes_2019}. 

     \begin{figure}[ht] 
        \centerline{\hspace*{0.015\paperwidth}
               \includegraphics[width=.65\paperwidth,clip=]{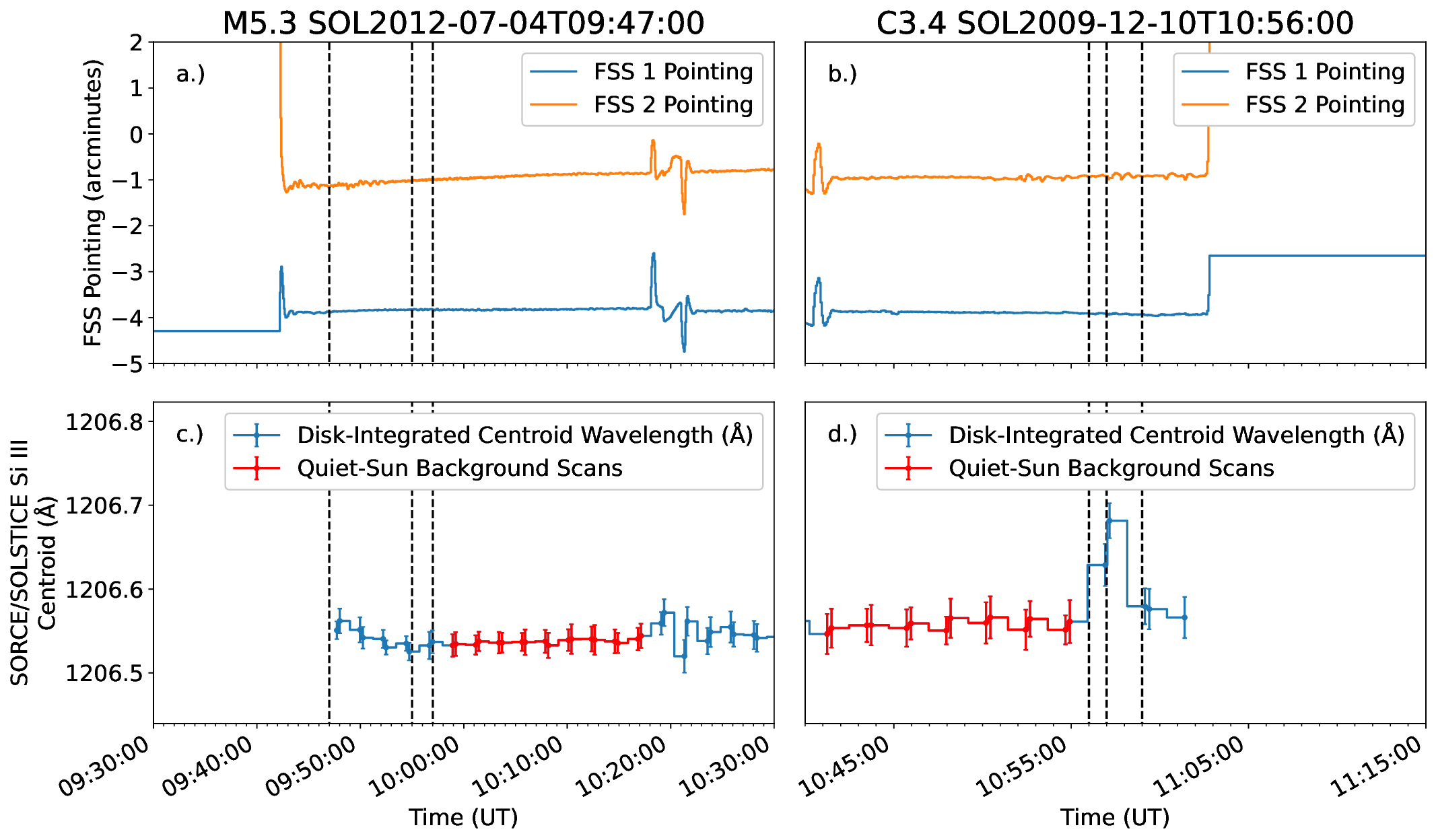}
              }
    \caption{FSS pointing information for an M5.3 and C3.4 flares, shown in panels a.) and b.), respectively. Panels c.) and d.) show the evolution of disk-integrated Si~\textsc{iii} centroid position during the respective events. The GOES flare start, peak, and end times are marked by black dashed lines, for the respective events.}
    \label{figure5.075}
    \end{figure}
    
    RHESSI provided observations of Hard X-ray (HXR) emission during flares. Comparing such emissions to enhancements in chromospheric and transition-region lines allows the evaluation of whether the observed line enhancements are likely driven by the injection of nonthermal electrons into the chromosphere, which produces such HXR emission via thick-target bremsstrahlung processes. In the absence of RHESSI observations, SXR observations from the $1-8\,\textrm{Å}$ channel of GOES/XRS provide context on whether Si~\textsc{iii} line enhancement and shifts were driven by impulsive or gradual phase heating. This requires the assumption of the Neupert effect, under which the derivative of $1-8\,\textrm{Å}$ emission is a proxy for the timing of HXR emission \citep{Neupert_1968}.
    
    For this study, a list of 129 flares of C-class or greater was generated, each event covered by SORCE/SOLSTICE wavelength calibration scans during at least the GOES peak time. From these events, a subset of 37 flares, in which the integrated Si~\textsc{iii} irradiance (between $1203.5$ and $1209.5\,\textrm{Å}$) increased by more than $10\%$ relative to the first scan during at least one scan in the observation, was selected. For each of the 37 events, the centroid position of the disk-integrated Si~\textsc{iii} line was determined via a preliminary Gaussian fitting. From this, the evolution of centroid position was manually inspected and compared to FSS observations, and shifts in pointing due to spacecraft manoeuvres were identified. Examples of these shifts are shown in Figure \ref{figure5.075}, manifesting as $\sim1^{\prime}$ deviations in pointing. The sample was further restricted to 14 flares that had no spacecraft manoeuvres during the GOES flare duration while SOLSTICE was taking observations, ensuring a consistent wavelength grid, and had at least three scans immediately before or after the flare that were uninterrupted by these manoeuvres to provide a quiet-Sun background profile. Finally, a set of 11 events with a signal-to-noise ratio ($\sigma$) greater than $2.5$ in Si~\textsc{iii} enhancement relative to quiet-Sun emission during at least one scan was chosen, using the uncertainties quoted in the SOLSTICE data product. This further ensured that any observed shifts were flare-related.
    
\begin{figure}[ht] 
        \centerline{\hspace*{0.015\paperwidth}
               \includegraphics[width=.65\paperwidth,clip=]{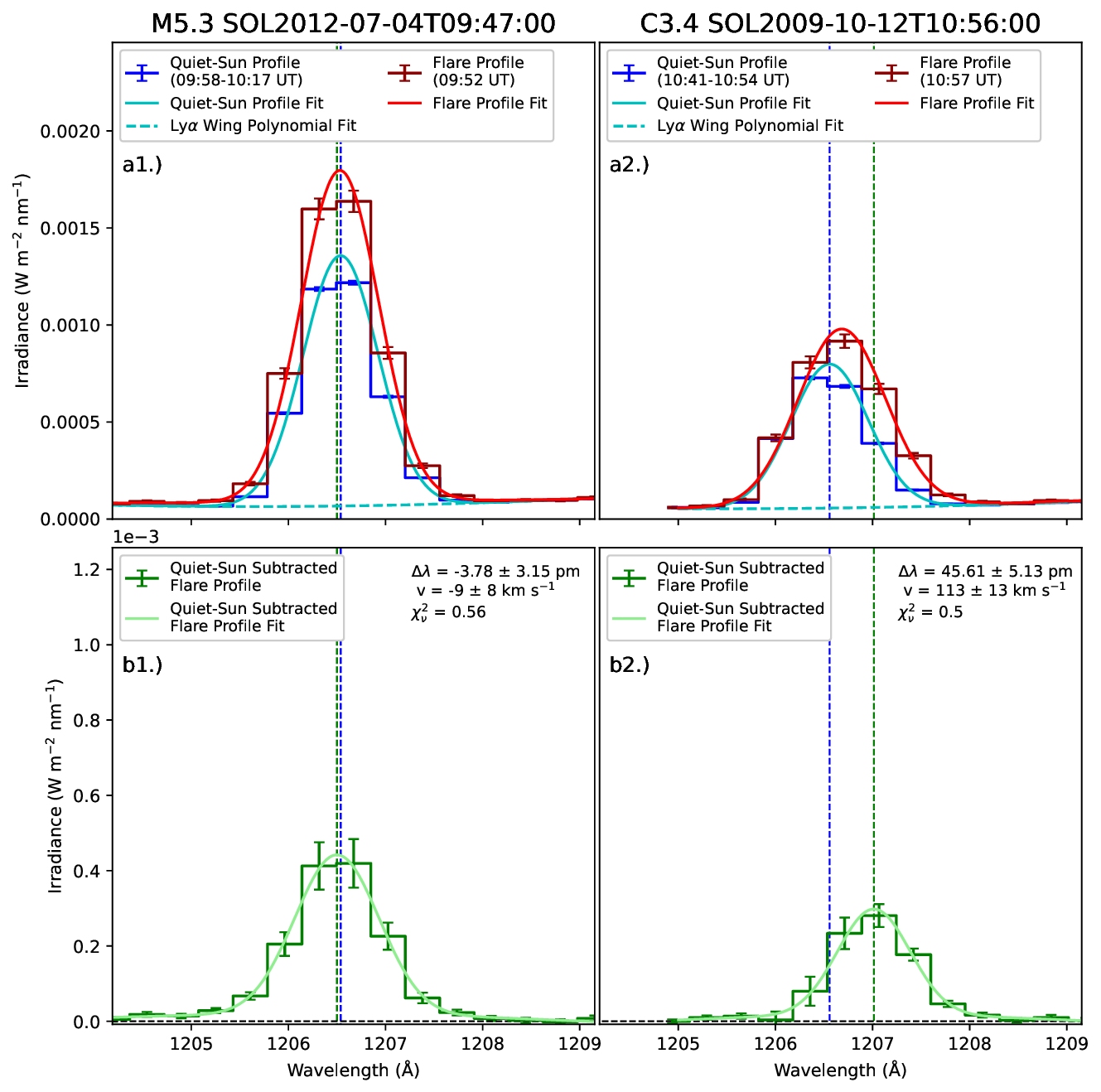}
              }
    \caption{ Observed and best-fit profiles to the Si~\textsc{iii} line for M5.3 (left) and C3.4 (right) flares. Panels a1.) \&  a2.) show flaring (maroon) and quiet-Sun (blue) profiles and their best fits (red and cyan, respectively). A dashed cyan line illustrates the polynomial fit to \Ly wing emission for the quiet-Sun profiles. Panels b1.) and b2.) show the associated quiet-Sun-subtracted flare profiles (dark green) and their respective best-fit profiles (light green) for each event. Dashed vertical lines indicate the centroid position for each event's quiet-Sun-subtracted profile (green) and quiet-Sun background profile (blue).}
    \label{figure5.2}
    \end{figure}

    For each of the 11 events, both a quiet-Sun and a flaring profile were calculated. The quiet-Sun profiles were calculated as an average of three or more scans before or after the GOES flare period of each event, while the flare profiles were defined as the profile that exhibited the greatest irradiance enhancement relative to quiet-Sun levels. Due to substantial uncertainties in irradiance, the disk-integrated nature of the observations and the relatively slow cadence of SOLSTICE observations, the time evolution of these profiles was not investigated. The quiet-Sun profiles were interpolated using a cubic-spline method and rebinned to match the wavelength grid of their respective flaring profiles. This was done to correct for a slow drift in the wavelength grid at which the SOLSTICE irradiance measurements were taken between scans. These drift-corrected quiet-Sun profiles were then subtracted from the disk-integrated flaring profiles for each event, providing a profile reflecting only flare-driven emission. Both the quiet-Sun and quiet-Sun-subtracted flare profiles were fit with a Gaussian and second-order polynomial for the Si~\textsc{iii} line and \Ly wing emission, respectively, using the \texttt{curve\_fit} routine in SciPy \citep{Virtanen_2020}. This provided both quiet-Sun and Doppler-shifted Si~\textsc{iii} wavelengths for each event. These fits were performed over the wavelength range $1203.5-1209.5\,\textrm{Å}$ for each scan, as illustrated in Figure \ref{figure5.2}. Uncertainty for each wavelength was determined as the square root of the covariance provided by \texttt{curve\_fit}. Doppler velocities were then calculated using the classical Doppler formula for a stationary observer:
    
    \begin{equation}  
    \label{equation1}
    v_{src} = c\times\frac{\Delta\lambda}{\lambda_0}
    \end{equation}
    \noindent 
    where $v_{src}$ is the velocity of the flaring source, $c$ is the speed of light, $\Delta\lambda$ is the Doppler shift, and $\lambda_0$ is the rest wavelength taken from the quiet-Sun profile's fit. In addition to the calculation of Doppler velocity, the integrated enhancement in irradiance over that of the quiet-Sun profile was calculated for Si~\textsc{iii}, providing context on the relative enhancement of the flaring profile for each event.
    
    \comment{}The calculated Doppler velocities were then analysed for any potential trend with solar longitude. Such trends have previously been identified in Sun-as-a-star observations of active regions by SDO/EVE, with Doppler speeds of up to $\sim100\,\textrm{km\,s}^{-1}$ being inferred near the solar limb, showing blueshifts and redshifts for eastern and western longitudes, respectively \citep{Hudson_2022,Fitzpatrick_2023}. The authors noted that this phenomenon was not seen in spatially-resolved observations of an active region by the EUV Imaging Spectrometer on Hinode \citep[Hinode/EIS;][]{Culhane_2007}, though they did not report any instrumental effects likely to explain the behaviour observed by EVE \citep{Bryans_2010}. \citet{Rajhans_2023} reported further spatially-resolved spectra of active regions from IRIS, and found no evidence of prograde flows. Spatially-integrated observations from Yohkoh's Bragg Crystal Spectrometer \citep[Yohkoh/BCS;][] {Culhane_1991} of active regions also show a different trend to that seen in EVE data, as discussed by \citet{Montgomery_prep}. The authors proposed three possible explanations for this disparity: (1) the trend observed in EVE data is an instrumental artifact, (2) the phenomenon does not manifest in the higher-temperature lines observed by BCS, or (3) BCS is insensitive to quiescent active-region emission and instead primarily detects flare-related emission, in which such flows may not exist. However, \citet{Woods_2025} reported similar longitude dependent Doppler shifts during flares to those previously identified in active regions. This suggests the observed active-region flows may contribute to observed flare spectral variability, though the underlying physics is not yet well understood. Should the effect be instrumentally driven, the relationship between longitude and shift for SOLSTICE may be different to that in EVE or BCS data, due to differences in instrumental design. 
    
\begin{figure}[ht!] 
        \centerline{
               \includegraphics[width=.55\paperwidth,clip=]{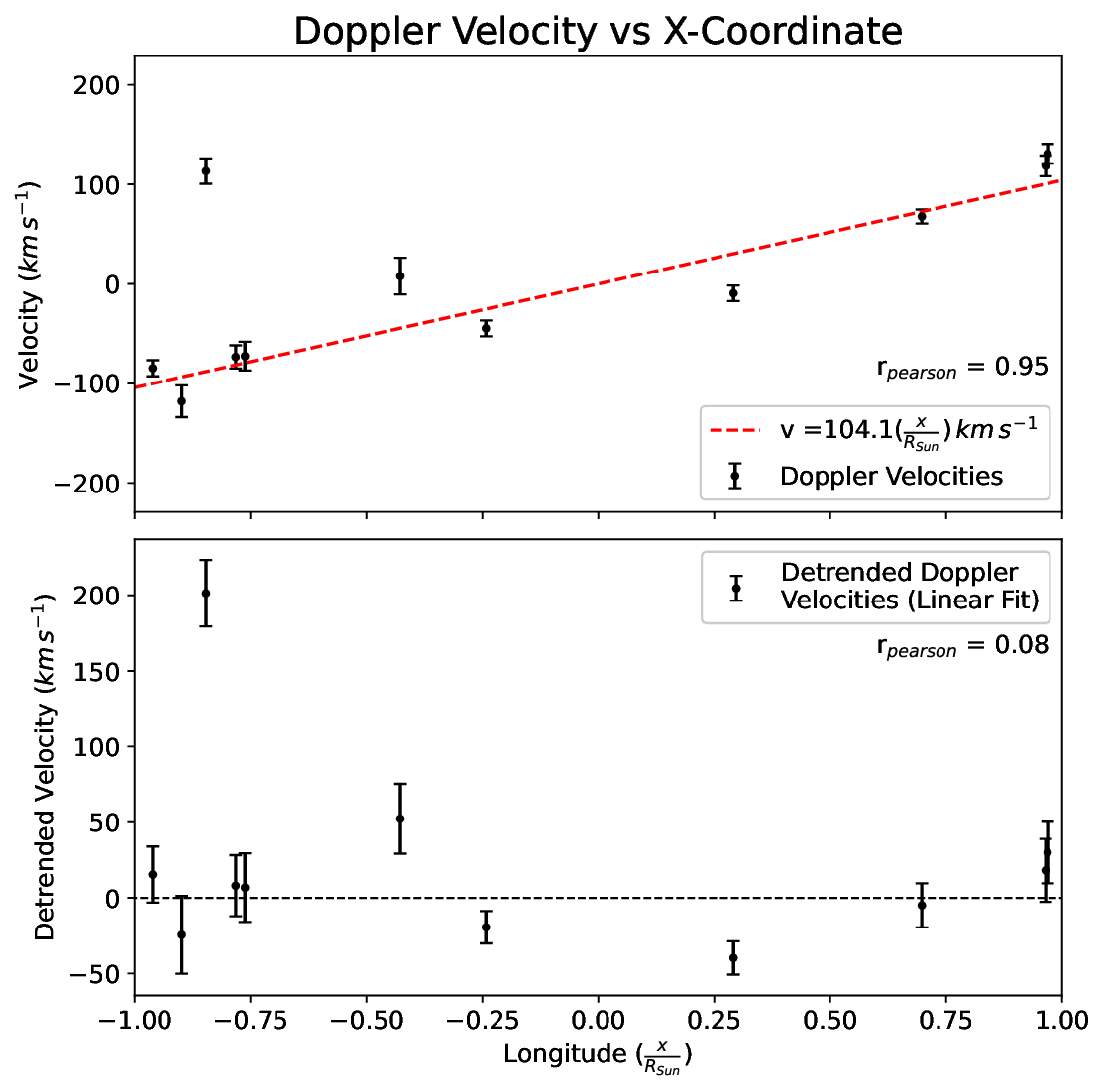}
              }
    \caption{Top panel: Doppler velocities, from fit Si~\textsc{iii} profiles, against longitude (in fraction of solar radius), data points shown as black dots, best fit line shown as dashed red line. Bottom panel: detrended Doppler velocities against longitude, data points shown as black dots, with dashed black line at zero velocity.}
    \label{figure5.4}
    \end{figure}

    In order to identify any trend in Doppler velocity with longitude in the SOLSTICE sample, longitudes for each event were acquired from the GOES event list\footnote{https://umbra.nascom.nasa.gov/goes/fits/} using the \texttt{get\_gev} routine with the GOES object in SolarSoft. For events with no location information in the GOES event list, the RHESSI flare list\footnote{https://hesperia.gsfc.nasa.gov/hessidata/dbase/hessi\_flare\_list.txt} and Heliophysics Event Knowledgebase \citep[HEK;][]{Hurlburt_2012} were queried for flare longitudes. The top panel of Figure \ref{figure5.4} presents each event's quiet-Sun-subtracted flare Doppler velocity against helioprojective $x$ coordinate as a fraction of apparent solar radius in arcseconds \citep{Thompson_2006}. A clear positive trend is seen with a Pearson correlation coefficient of $0.95$, excluding an outlier near the Eastern limb that was redshifted, suggesting similar `prograde flows' to those described before for EVE observations. To retrieve signatures of flare motions from these data, a linear fit to the data was subtracted, providing the detrended velocities displayed in the bottom panel of Figure \ref{figure5.4}. It should be noted that if the sample is dominated by either condensation or evaporation, the detrending may have overcorrected the data, leading to a respective under- and overestimation of the over- and under-represented flows' magnitudes. Furthermore, past ray-trace simulation results show SOLSTICE to exhibit systematic wavelength shifts due to lack of consistent alignment of the entrance slit with the solar North-South line, which may introduce unaccounted for uncertainty in measured Doppler velocity (Woods, T.N., Personal Communication, 2025).\commentend{}
    
    For each of the events in the sample, the relative timing of peak Si~\textsc{iii} enhancement and the respective Doppler shift at this time was compared to the timing of enhancement in $1-8\,\textrm{Å}$ emission. This enabled the assessment of whether the line enhancements and corresponding Doppler shifts were driven by impulsive heating by nonthermal electrons or whether a process during the gradual phase, such as thermal conduction to the chromosphere, was more likely responsible. \citep[e.g.][]{Zarro_1988}. Events that were observed by EUVS-E, RHESSI and AIA were analysed in greater detail. EUVS-E provides further context on the timing of enhancement with its superior cadence to SOLSTICE, further clarifying which flare processes likely drove Si~\textsc{iii} enhancement and line shifts. RHESSI allows direct comparison of the timing of Si~\textsc{iii} emission to enhancement in HXR emission between $25-50\,\textrm{keV}$ that probes heating of the chromosphere by nonthermal electrons. Imaging from the $304\,\textrm{Å}$ and $1600\,\textrm{Å}$ channels of AIA allowed for the identification of any eruptions or jets potentially responsible for line shifts. 
   
    \begin{figure}[ht!] 
        \centerline{\hspace*{0.015\paperwidth}
               \includegraphics[width=.55\paperwidth,clip=]{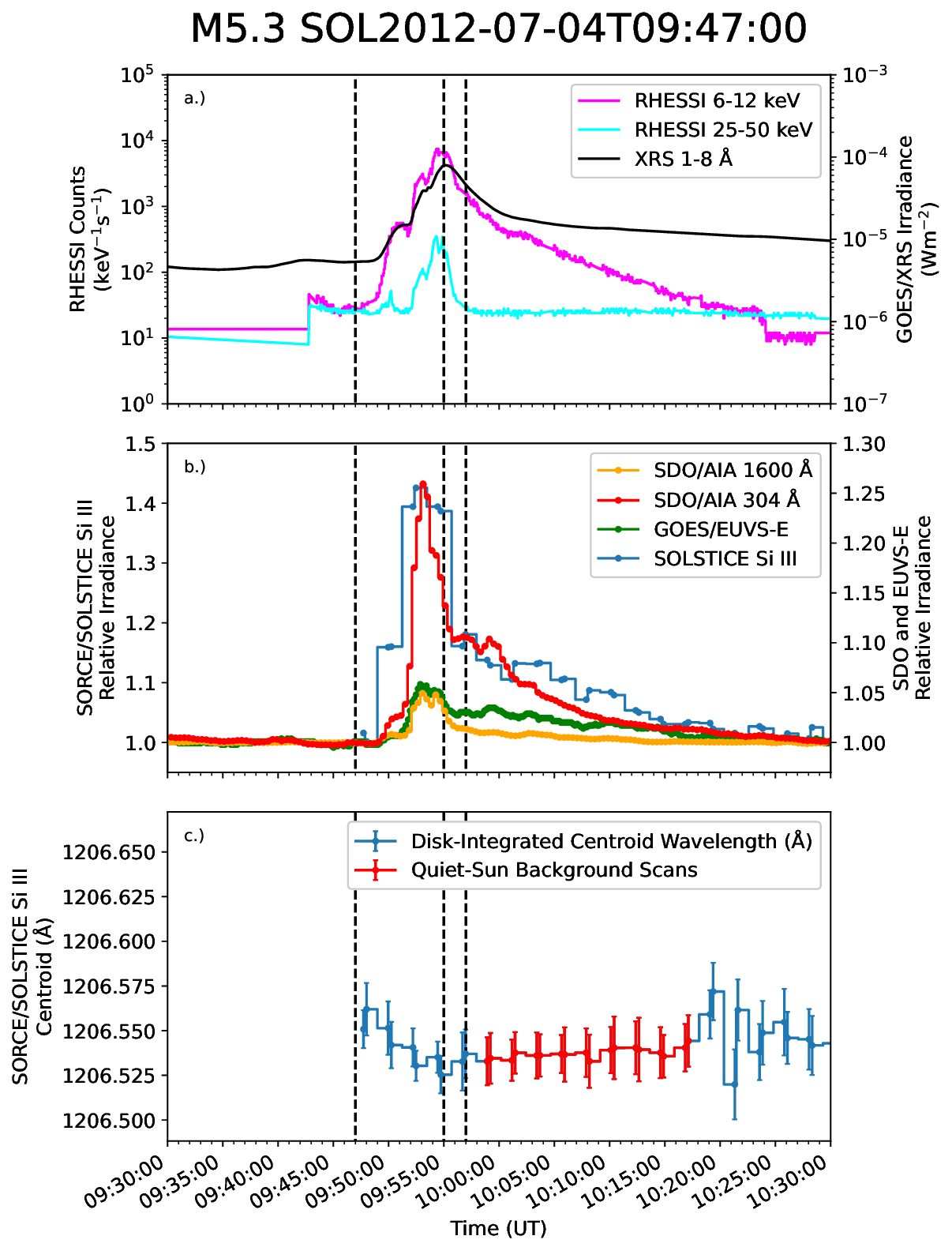}
               \hspace*{-0.03\textwidth}
              }
    \caption{ Lightcurves of M5.3 flare. Panel a.) shows X-ray observations in $1-8\,\textrm{Å}$ (black), $6-12\,\textrm{keV}$ (magenta) and $25-50\,\textrm{keV}$ (cyan). Panel b.) shows chromospheric emission in a variety of lines, including the $1600\,\textrm{Å}$ (orange) and $304\,\textrm{Å}$ (red) channels of SDO/AIA, the \Ly line from GOES/EUVS-E (green) and the Si~\textsc{iii} line from SORCE/SOLSTICE (blue). Panel c.) shows the centroid position of the Si~\textsc{iii} line during the event in blue. The scans that were averaged to generate quiet-Sun profiles are highlighted in red. In each panel, the GOES flare start, peak, and end times are marked by black dashed lines in order from left to right.}
    \label{figure5.1}
    \end{figure}

\section{Results}
    \label{S- Results} 
     Figure \ref{figure5.1} shows X-ray and UV lightcurves along with disk-integrated Si~\textsc{iii} centroid position for the M5.3 flare \flrthree. Panel a.) of Figure \ref{figure5.1} shows HXR and SXR observations from RHESSI and GOES/XRS, respectively. Comparing these data with chromospheric emissions in panel b.), it is seen that Si~\textsc{iii} (09:52:29 UT), along with $1600\,\textrm{Å}$ (09:53:04 UT), $304\,\textrm{Å}$ (09:53:07 UT) and \Ly enhancement (09:52:52 UT), peaks closely in time to a burst in $25-50\,\textrm{keV}$ emission around 09:53:05 UT. This suggests that the enhancement was predominantly driven by the injection of nonthermal energy into the chromosphere by accelerated electrons.

    \begin{figure}[ht!] 
        \centerline{
               \includegraphics[width=.75\paperwidth,clip=]{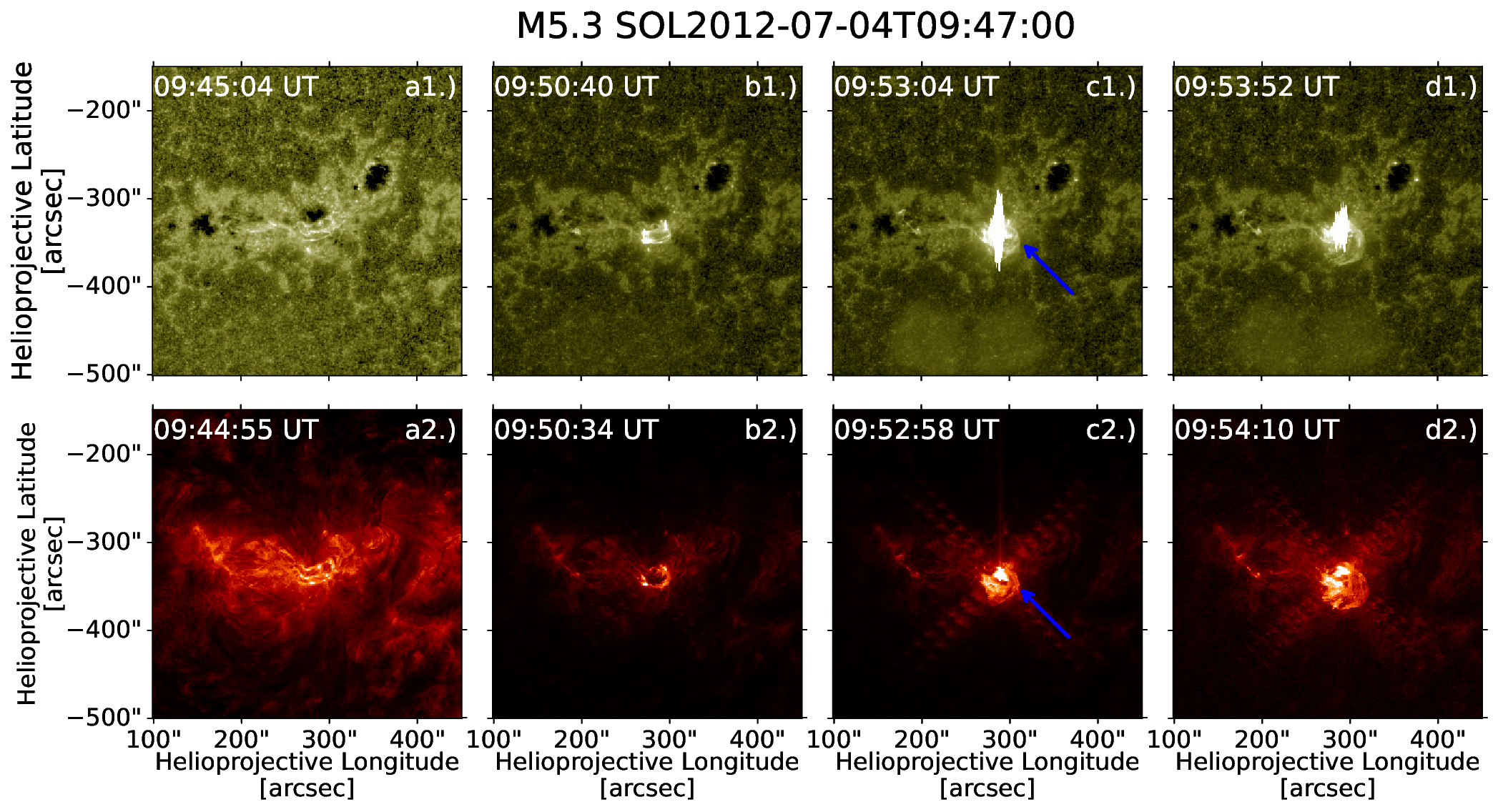}
              }
    \caption{ SDO/AIA images of M5.3 flare in the $1600\;\textrm{Å}$ (top) and $304\;\textrm{Å}$ (bottom) channels. Panels a1.) and a2.) show the active region (NOAA 11515) before the flare. b1.) and b2.) show flare footpoints during the impulsive phase. A blue arrow points to a bright eruption in panels c1.) and c2.). Panels d1.) and d2.) show the active region after the GOES end time of the flare}
    \label{figure5.3}
    \end{figure}
    
     The fitted line profiles for the M5.3 flare are shown in the left panels of Figure \ref{figure5.2}. The top left panel displays the quiet-Sun and unsubtracted flare profiles along with their corresponding fits, while the bottom left panel shows the quiet-Sun-subtracted flare profile and its fit. The subtracted flare profile of the M5.3 event shows a wavelength shift of $\Delta\lambda = -3.78\pm3.15\,\textrm{pm}$ relative to the quiet-Sun profile, corresponding to a weakly blueshifted Doppler velocity of $-9\pm8\,\textrm{km\,s}^{-1}$. Applying a systematic correction (detailed further at the end of this section) for longitude-based Doppler shift, provides a faster blueshifted velocity of $-39.75\pm11.00\,\textrm{km\,s}^{-1}$. Context images from SDO/AIA for the M5.3 flare are shown in Figure \ref{figure5.3}, panels a1.) \& a2.) show the preflare active region (NOAA 11389) in the $1600\,\textrm{Å}$ and $304\,\textrm{Å}$ channels, respectively. Flare footpoints can be seen in the impulsive-phase emission shown in panels b1.) \& b2.). Panels c1.) \& c2.) show an eruption that occurs at a similar time (09:53:04 \& 09:52:58 UT) to when the Si~\textsc{iii} peak enhancement is seen (09:52:29 UT), with the line exhibiting a blueshift at this time. It is hence possible that this blueshift was driven by Doppler-shifted emission from this bright erupting material \citep{RubiodaCosta_2009,Wauters_2022}. This scenario was similarly suggested to explain a blue asymmetry observed in the \Ly line during the same M5.3 event in \citet{Majury_2025}.

     \begin{figure}[ht!] 
        \centerline{\hspace*{0.015\paperwidth}
               \hspace*{-0.03\textwidth}
               \includegraphics[width=.55\paperwidth,clip=]{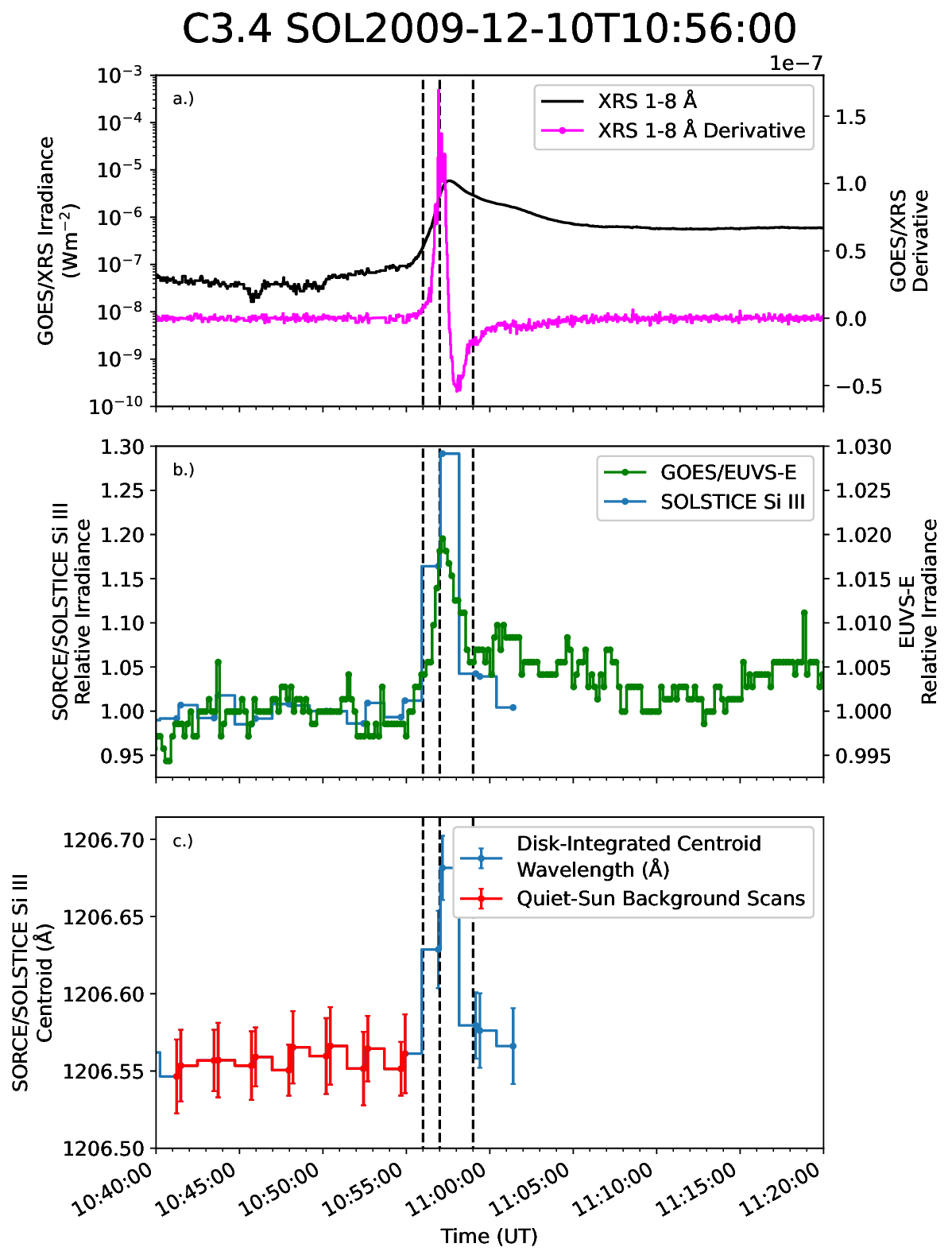}
              }
    \caption{Lightcurves of C3.4 flare. Panel a.) displays $1-8\,\textrm{Å}$ observations (black). Panel b.) shows Si~\textsc{iii} (blue) and \Ly (green) observations from SORCE/SOLSTICE and GOES/EUVS-E, respectively. Panel c.) shows the centroid position of the Si~\textsc{iii} line during the event in blue. The scans that were averaged to generate quiet-Sun profiles are highlighted in red. In each panel, the GOES flare start, peak, and end times are marked by black dashed lines in order from left to right.}
    \label{figure5.15}
    \end{figure}
    
    Figure \ref{figure5.15} shows X-ray and UV lightcurves along with disk-integrated Si~\textsc{iii} centroid position for the C3.4 flare \flrfour. Panel a.) of Figure \ref{figure5.15} shows SXR emission during the C3.4 event. Comparing this with Si~\textsc{iii} and \Ly emission, seen in panel b.), shows that these emissions correlate more closely with gradual SXR emission than in the M5.3 event. Although no RHESSI data were available during the C3.4 event for direct comparison with the event's HXR emission, the Si~\textsc{iii} line peaks at 10:57:10 UT, closer in time to the $1-8\,\textrm{Å}$ flux-derivative peak at 10:57:06 UT than the peak in $1-8\,\textrm{Å}$ flux at 10:57:22 UT (Figure \ref{figure5.15}). Under the assumption that the Neupert effect holds for this event, this suggests the Si~\textsc{iii} enhancement, similarly to in the M5.3, was impulsively driven by nonthermal heating of the chromosphere \citep{Neupert_1968}. While this comparison is somewhat limited by the low cadence of SORCE/SOLSTICE, similarly chromospheric \Ly emission observed at a higher cadence shows a similar trend, peaking at 10:57:11 UT, adding validity to the Si~\textsc{iii} line being impulsively enhanced for this event. Panel c.) shows a redshift in the disk-integrated Si~\textsc{iii} line cotemporally with the line's enhancement. The fitted line profiles for the C3.4 flare are shown in the right panels of Figure \ref{figure5.2}. The top right panel displays the quiet-Sun and unsubtracted flare profiles along with their corresponding fits, while the bottom right panel shows the quiet-Sun-subtracted flare profile and its respective fit. The subtracted flare profile of the C3.4 event exhibits a positive wavelength shift of $45.61\pm5.13\,\textrm{pm}$, indicating a Doppler velocity of $113\pm13\,\textrm{km\,s}^{-1}$. After applying the same longitude correction as with the M5.3 event, this velocity becomes an even larger $201.36\pm21.94\,\textrm{km\,s}^{-1}$.

    The detrended Si~\textsc{iii} Doppler velocities calculated for each of the 11 events in the sample are tabulated in Table \ref{table5.1}. Events with a shift with a signal-to-noise of three or more ($\geq3\sigma$) are denoted as having a blueshift or redshift, with events below this threshold being marked with `None' as the measured shift was below significance. Only the aforementioned M5.3 and C3.4 events displayed a significant Doppler velocity, with a further three events having a shift above one signal-to-noise. Intriguingly, the event showing a fast redshift, indicative of condensation, had a much smaller GOES class of C3.4 than the M5.3 event, which showed a blueshift, indicating evaporation. It is possible that condensation and/or evaporation occurred in the other nine events, but were not detected due to large uncertainties. For the quiet-Sun-subtracted flare profile fits, reduced chi-squared values were below unity for 10 of the events. This indicates good fits, considering the uncertainties in irradiance, with the X3.6 event exhibiting the largest value of $\chi^2_\nu = 11.42$. The peak flare enhancement over the quiet-Sun background for each event is tabulated in Table \ref{table5.2}, with the greatest enhancement being $175.77\pm9.74\%$ during the X3.6 event, and the smallest enhancement being $13.75\pm4.83\%$ during the C1.5 event. The Pearson correlation coefficient between $1-8\,\textrm{Å}$ flux and Si~\textsc{iii} percentage enhancement is $0.92$, showing a strong relationship. The signal-to-noise ratio in enhancement for Si~\textsc{iii} was greater than three for 10 of the 11 events. Peak Si~\textsc{iii} enhancement occurred before the peak in $1-8\,\textrm{Å}$ flux for all 11 events, suggesting the Si~\textsc{iii} enhancement and corresponding line shifts were driven by an impulsive process, such as heating by nonthermal electrons, for each event.

\section{Discussion \& Conclusions}
    \label{S- Discussion}
    This study presents an analysis of the flaring Si~\textsc{iii} profile using SORCE/SOLSTICE observations, revealing a potential new diagnostic of flare-associated mass motions in the chromosphere and transition region. In analysing 11 X-, M- and C-class flares, we find that two of the flares exhibited significant Doppler velocities, at above the $3\sigma$ level. The detection of mass motions was likely limited by large instrumental uncertainties, on the order of tens of $\textrm{km\,s}^{-1}$, as lines at similar transition region temperatures to Si~\textsc{iii} typically only exhibit velocities of a few tens of $\textrm{km\,s}^{-1}$ for both red- and blueshifted emission during flares \citep{Milligan_2006_1,Milligan_2006_2,Ding_2019}.
      
     The blueshift velocity of $-39.75\pm11.00\,\textrm{km\,s}^{-1}$ exhibited during the M5.3 event is somewhat fast for gentle evaporation, with similarly fast velocities of $\sim50\,\textrm{km\,s}^{-1}$ previously being seen in the $\textrm{Ly}\epsilon$ line in observations from the Multiple EUV Grating Spectrograph (MEGS) on SDO/EVE, which was attributed to filament eruption associated with the event \citep{Brown_2016}. Additionally, spatially-resolved observations of Si~\textsc{iv} ($1403\,\textrm{Å}$) found blueshifted Doppler velocities of up to $-120\;\textrm{km}\,\textrm{s}^{-1}$ in a jet associated with a flare, although the relative contribution of jet emission to the overall flare excess is not discussed \citep{Zhang_2021}. Previous observations of a flare-related filament eruption in chromospheric \Ly images from the Transition Region and Coronal Explorer \citep[TRACE;][]{Handy_1999} found the projected velocity of the filament to be $300\pm50\,\textrm{km}\,\textrm{s}^{-1}$, while the surface intensity ($\textrm{erg}\,\textrm{cm}^{-2}\,\textrm{s}^{-1}$) of the filament was $\sim\frac{1}{10}$ that of the flare ribbons, the filament surface area was considerably larger \citep{RubiodaCosta_2009}.  Inspection of the AIA images in Figure \ref{figure5.3} reveals an eruption that travels approximately $10^{\prime\prime}$ over $72\,\textrm{s}$ towards south-east between panel b2.) and c2.). Assuming the eruption moves at a similar or faster speed perpendicular to the solar surface, at $\sim100\,\textrm{km\,s}^{-1}$, infers a projected line-of-sight velocity of $\sim90\,\textrm{km\,s}^{-1}$. While this is more than fast enough to explain the observed blueshift, it is not clear how much emission from the filament contributed to the flare signal, thus the relative contributions of evaporation and eruption motion to the observed shift are not clear. It is also noted that the C8.4 event in the sample was also associated with an eruption, but did not show a significant Doppler shift.
    
    \begin{table}[ht!]
    \begin{tabular}{cccccc}
    \hline
    \textbf{Class} & \shortstack{\textbf{GOES Flare}\\\textbf{Peak (UT)}} & \textbf{\shortstack{Detrended\\Velocity ($\textrm{km\,s}^{-1}$)}} & $\chi^2_\nu$ & \textbf{$\sigma$ (Shift Type)} & \textbf{\shortstack{Stonyhurst\\Coordinate ($^{\circ}$)}}\\
    \hline
    X3.6 & 2005-09-09T09:59:00 & -24.44 $\pm$ 25.70 & 11.415 & 0.95 (None) & S11E66 \\
    M8.3 & 2004-01-07T10:27:00 & 15.40 $\pm$ 18.53 & 0.742 & 0.83 (None) & N06E75 \\
    M5.3 & 2012-07-04T09:55:00 & -39.75 $\pm$ 11.00 & 0.560 & 3.61 (Blueshift) & S20W18 \\
    M2.7 & 2005-01-19T10:24:00 & -4.95 $\pm$ 14.59 & 0.477 & 0.34 (None) & N18W47 \\
    M1.5 & 2003-03-20T11:31:00 & 18.13 $\pm$ 20.84 & 0.539 & 0.87 (None) & N04W75 \\
    M1.3 & 2005-05-06T11:28:00 & 29.94 $\pm$ 20.36 & 0.190 & 1.47 (None) & N04W76 \\
    C8.4 & 2011-12-30T10:32:00 & 8.04 $\pm$ 20.15 & 0.602 & 0.40 (None) & N23E58 \\
    C7.6 & 2005-09-10T10:28:00 & 6.77 $\pm$ 22.65 & 0.389 & 0.30 (None) & S12E51 \\
    C3.4 & 2009-12-10T10:57:00 & 201.36 $\pm$ 21.94 & 0.504 & 9.18 (Redshift) & N20E64 \\
    C3.2 & 2004-08-18T10:44:00 & -19.42 $\pm$ 10.67 & 0.341 & 1.82 (None) & N03E14 \\
    C1.5 & 2010-03-27T10:14:00 & 52.26 $\pm$ 23.08 & 0.912 & 2.26 (None) & N14E26 \\
    \hline
    \end{tabular}
    \caption{ Doppler velocities for the quiet-Sun-subtracted Si~\textsc{iii} profile during each flare, GOES peak times, reduced chi-squared values of quiet-Sun-subtracted flare profile fits, signal-to-noise ratio with shift type (blueshift/redshift/none) based on whether velocity had $\geq3\sigma$, along with each flare's Stonyhurst longitude and latitude.}
    \label{table5.1}
    \end{table}

    A remarkably fast redshift velocity was seen in the C3.4 event at $201.36\pm21.94\,\textrm{km\,s}^{-1}$. Such a fast redshift velocity has seldom been observed before, with some of the fastest redshifts of $\sim100\,\textrm{km\,s}^{-1}$ having been inferred during flares in the chromospheric $\textrm{H}\alpha$ and Si~\textsc{iv} ($1403\,\textrm{Å}$) lines \citep{Ichimoto_1984,Tian_2015}. \citet{Simoes_2015} reported a Doppler velocity up to $250\,\textrm{km\,s}^{-1}$ in the coronal Fe~\textsc{XII} ($\textrm{T} = 10^{6.2}$) line using observations from Hinode/EIS. While this velocity is comparable to the $\sim200\,\textrm{km\,s}^{-1}$ seen in the Si~\textsc{iii} line, there was also a stationary component in the Fe~\textsc{XII} line, with no clear secondary component seen in Si~\textsc{iii}. It is also noted that the Si~\textsc{iii} line forms at a lower temperature than Fe~\textsc{XII}, with higher density making its high downflow velocity more unusual. Theoretical work suggests that such a high velocity would require either an abnormally high energy flux or an atypically under-dense pre-flare chromosphere \citep{Fisher_1987, Longcope_2014}. Such low chromospheric densities may be driven by mass ejections in jets or filament eruptions that occur before the flare onset \citep{Shimojo_2000,Feng_2012,Filipov_2021,Saqri_2023}. Another physical process, such as coronal rain, where downflows of over $100\,\textrm{km\,s}^{-1}$ have been observed in the chromospheric temperature H$\alpha$ and He~\textsc{ii} lines, may also drive the observed Si~\textsc{iii} velocities \citep{Vashalomidze_2015,Antolin_2015}. However, given that the phenomenon is much fainter than flare footpoints and generally occurs after the impulsive phase, during which the Si~\textsc{iii} velocities were observed, this seems unlikely. 
    
    The detection of such high velocities, particularly during the C3.4 event, in these disk-integrated SORCE/SOLSTICE observations is notable, as the quiet-Sun-subtracted flare signal represents the entire flaring region, suggesting these high velocities do not simply represent an extreme flare kernel. It is possible that the $\sim10\,\textrm{s}$ raster time across the line, combined with short-term variation in flare enhancement, could result in systematically driven shifts. However, as shown in panel c.) in both Figure \ref{figure5.1} and Figure \ref{figure5.15}, the respective centroid positions of the disk-integrated Si~\textsc{iii} profiles do not indicate a change in shift direction between raster scans associated with Si~\textsc{iii} enhancement. This suggests rastering effects do not play a substantial role in the observed shifts. \comment{}The large shift during the C3.4 event may also possibly be explained by strong enhancement of a line blend in the Si~\textsc{iii} red wing during the event. However, the CHIANTI line list does not predict any lines within $\pm4\,\textrm{Å}$ of the Si~\textsc{iii} line centroid, thus any apparent shifts due to blends are expected to be much larger ($>1000\,\textrm{km\,s}^{-1}$) than those observed. However, any uncatalogued line blends closer than $4\,\textrm{Å}$ may be difficult to detect given the modest spectral resolution of SOLSTICE. While a substantial trend in measured Doppler velocity with solar longitude was corrected for, our assumption of a first-order polynomial fit to detrend this data may not be comprehensive. A more complete correction for this trend may be achieved via ray-trace simulations using the SOLSTICE optical design, which may reveal more modest velocities. Such analysis would also further elucidate whether this observed `prograde flow' behaviour is likely instrumentally driven or due to a physical process associated with flares\commentend{}. The $\chi^2_\nu$ values for all but one of the fitted quiet-Sun-subtracted flare profiles were less than unity, demonstrating that the single Gaussian fit was sufficient given the irradiance uncertainty and wavelength sampling of the SOLSTICE data. However, for the X3.6 event, which had a large chi-squared value, a better fit may have been achieved with two Gaussians, representing a stationary and moving source. Such two-component profiles have previously been observed during flares in transition-region and chromospheric lines such as the Mg h \& k lines \citep{Graham_2020}.

    \begin{table}[ht!]
    \begin{tabular}{ccccccccc}
    \hline
    \textbf{Class} & \textbf{Date} & 
    \shortstack{\textbf{GOES Flare}\\\textbf{Start (UT)}} & 
    \shortstack{\textbf{GOES Flare}\\\textbf{Peak (UT)}} & 
    \shortstack{\textbf{Si\,\textsc{iii}}\\\textbf{Peak (UT)}} & 
    \shortstack{\textbf{Si\,\textsc{iii}}\\\textbf{Enh. (\%)}} & 
    \shortstack{\textbf{$\sigma$}\\\textbf{(Si\,\textsc{iii})}} \\
    \hline
    X3.6 & 2005-09-09 & 09:42:00 & 09:59:00 & 09:56:46 & 175.77 $\pm$ 9.74 & 18.04 \\
    M8.3 & 2004-01-07 & 10:14:00 & 10:27:00 & 10:23:05 & 74.87 $\pm$ 5.97 & 12.54 \\
    M5.3 & 2012-07-04 & 09:47:00 & 09:55:00 & 09:52:29 & 31.33 $\pm$ 5.04 & 6.22 \\
    M2.7 & 2005-01-19 & 10:19:00 & 10:24:00 & 10:20:58 & 74.82 $\pm$ 7.51 & 9.96 \\
    M1.5 & 2003-03-20 & 11:25:00 & 11:31:00 & 11:28:07 & 66.05 $\pm$ 6.76 & 9.78 \\
    M1.3 & 2005-05-06 & 11:11:00 & 11:28:00 & 11:24:25 & 30.78 $\pm$ 5.38 & 5.72 \\
    C8.4 & 2011-12-30 & 10:27:00 & 10:32:00 & 10:30:32 & 22.84 $\pm$ 4.53 & 5.04 \\
    C7.6 & 2005-09-10 & 10:24:00 & 10:28:00 & 10:27:25 & 16.56 $\pm$ 4.49 & 3.69 \\
    C3.4 & 2009-12-10 & 10:56:00 & 10:57:00 & 10:57:09 & 29.16 $\pm$ 5.55 & 5.25 \\
    C3.2 & 2004-08-18 & 10:40:00 & 10:44:00 & 10:42:02 & 41.55 $\pm$ 5.15 & 8.06 \\
    C1.5 & 2010-03-27 & 10:05:00 & 10:14:00 & 10:12:06 & 13.75 $\pm$ 4.83 & 2.85 \\
    \hline
    \end{tabular}
    \caption{Si\,\textsc{iii} peak enhancements (\%) over quiet-Sun background during each of the 11 flares, with signal-to-noise ratios and Si\,\textsc{iii} peak times (UT).}
    \label{table5.2}
    \end{table}
    
    Further insight may be provided from comparisons between the existing SORCE/SOLSTICE observations and observations of other spectral lines during the flares by instruments such as the Coronal Diagnostic Spectrometer of the Solar and Heliospheric Observatory \citep[SOHO/CDS;][]{Harrison_1995}, Hinode/EIS, or SDO/EVE. Unfortunately, no observations from CDS or EIS were available for the events in our sample. Spectrally resolved observations of the He~\textsc{ii} $304\,\textrm{Å}$ line during the C8.4 event are available from SDO/EVE MEGS-A. However, as the correction required to He~\textsc{ii} wavelength shifts described in \citet{Chamberlin_2016} has been subject to recent scepticism, no comparisons of He~\textsc{ii} were drawn with Si~\textsc{iii} line shifts (Chamberlin, P.C., Personal Communication, 2025). Though a blueshift during the M5.3 event was attributed to an associated filament eruption, the lack of spatially-resolved spectra makes it impossible to conclusively determine the relative contribution of flare and eruption emission to the observed shift. A dedicated study using spectrally resolved imaging data from instruments such as EIS or IRIS may allow for these relative contributions to be determined for other events with associated eruptions, further clarifying the significance of eruptions in driving flare-associated line shifts . Analysis of the time evolution of Si~\textsc{iii} shift velocities may also yield further insight into flare dynamics. However, such analysis is challenging due to large uncertainties in the sample of events observed by SOLSTICE. These observations of the Si~\textsc{iii} line would also benefit from comparisons to the results of radiative hydrodynamics simulations. \citet{Kerr_2019} performed such simulations for the Si atom, but focused on the Si~\textsc{iv} line observed by IRIS. 

    The observed line shifts in the Si~\textsc{iii} line in this study should inform future studies into Doppler shifts of chromospheric and transition-region lines. Of particular interest is the upcoming EUV High-Throughput Spectroscopic Telescope \citep[SOLAR-C/EUVST;][]{Shimizu_2019}; its LW3 band (1115-1275 Å) will perform high-cadence ($1\,\textrm{s}$), high-resolution ($0.008\,\textrm{Å}$) imaging spectrometry measurements of the Si~\textsc{iii} line, surpassing the capabilities of the SORCE/SOLSTICE calibration scans. Additionally, high-cadence ($1\,\textrm{s}$) and resolution ($0.0033\,\textrm{Å}$) imaging spectroscopy of the Si~\textsc{iii} line will be provided by the upcoming Solar eruptioN Integral Field Spectrograph \citep[SNIFS;][]{Herde_2024} sounding rocket mission, which may fortuitously capture a flare. This should allow for more precise measurements of Si~\textsc{iii} line shifts during flares, with the faster cadence allowing for deeper comparisons with HXR emission measurements that probe nonthermal heating by instruments such as the Spectrometer Telescope for Imaging X-rays on Solar Orbiter \citep[SolO/STIX;][]{Muller_2020,Krucker_2020} and the Hard X-ray Imager on the Advanced Space-based Solar Observatory \citep[ASO-S/HXI;][]{Zhang_2019}. These observations may also benefit from comparisons with other chromospheric and transition-region lines observed by SOLAR-C and by other upcoming state-of-the-art instruments such as the Multi-slit Solar Explorer \citep[MUSE;][]{DePontieu_2020}. 

    The observations in this study demonstrate that the transition-region Si~\textsc{iii} line can exhibit Doppler shifts in either direction during flares, with potential explanations for the observed redshift lying in chromospheric condensation, and the observed blueshift potentially being driven by bright filament eruptions or jets. To the best of the authors' knowledge, this is the first publication to study Doppler shifts in the Si~\textsc{iii} line induced by flare processes. We reveal the line to be a potentially useful diagnostic of flare dynamics at the boundary between the chromosphere and transition region, with future observations of the line and other lines at similar temperatures, using contemporary and upcoming instruments, providing an opportunity for further insights into chromospheric condensation and eruptive flare dynamics to be yielded. However, our work also urges caution for future studies employing spatially-integrated spectral flare observations, as substantial systematic shifts due to varying flare longitude are seen in SOLSTICE data. 
    
\begin{acks}
L.H.M. acknowledges support from the Department for the Economy (DfE) Northern Ireland postgraduate studentship scheme. R.O.M. acknowledges support from STFC grant ST/X000923/1. The authors thank Prof. Martin Snow and Stéphane Béland for providing information surrounding and access to the SORCE/SOLSTICE FSS data. The authors additionally thank Dr. Phil Chamberlin for providing guidance on interpreting SDO/EVE data. We also thank the anonymous referee for insightful comments that helped significantly improve the manuscript.
\end{acks}

\bibliographystyle{spr-mp-sola}
\bibliography{bibliography.bib} 
\appendix
    
    
    
\section{Lyman-Alpha Enhancements in Sample}

    \begin{table}[ht]
        \begin{tabular}{cccccc}
        \hline
        \textbf{Class} & \textbf{Date} & 
        \textbf{Ly$\alpha$ Enh. (\%)} & 
        \textbf{$\sigma$} \\
        \hline
        X3.6 & 2005-09-09 & 3.79 $\pm$ 3.27 & 1.16 \\
        M8.3 & 2004-01-07 & 4.61 $\pm$ 3.04 & 1.52 \\
        M5.3 & 2012-07-04 & 0.60 $\pm$ 3.18 & 0.19 \\
        M2.7 & 2005-01-19 & 1.91 $\pm$ 4.00 & 0.48 \\
        M1.5 & 2003-03-20 & 3.56 $\pm$ 3.78 & 0.94 \\
        M1.3 & 2005-05-06 & 0.66 $\pm$ 3.48 & 0.19 \\
        C8.4 & 2011-12-30 & 0.71 $\pm$ 2.94 & 0.24 \\
        C7.6 & 2005-09-10 & 1.74 $\pm$ 3.16 & 0.55 \\
        C3.4 & 2009-12-10 & 1.01 $\pm$ 3.69 & 0.27 \\
        C3.2 & 2004-08-18 & 3.21 $\pm$ 3.15 & 1.02 \\
        C1.5 & 2010-03-27 & 1.41 $\pm$ 3.45 & 0.41 \\
        \hline
        \end{tabular}
        \caption{\Ly peak enhancements (\%) over quiet-Sun background during each of the 11 flares}
        \label{table5.3}
    \end{table}
    
    Percentage enhancements above quiet-Sun emission for the \Ly line ($1209.5-1222.0\,\textrm{Å}$) are presented in Table \ref{table5.3}, analogous to the Si~\textsc{iii} enhancements shown earlier in Table \ref{table5.2}. These \Ly percentage enhancements were comparably weaker than those observed in the Si~\textsc{iii} line, the largest being $4.61\%$ during the M8.3 event. For Si~\textsc{iii}, the smallest observed enhancement was $13.75\pm4.83\%$ during a C1.5 event. The signal-to-noise ratio in \Ly enhancement was less than unity for 8 of the 11 events, with the largest value being 1.52 during the M8.3 flare. Due to this low $\sigma$, measurement of Doppler velocities of the line was not pursued further. Furthermore, any measurement of Doppler velocity would be complicated by the varied formation heights of different parts of the line \citep{Vernazza_1973,Vernazza_1981,Fontenla_1991}.
    
\end{article} 
\end{document}